\documentclass[sigconf,natbib=false]{acmart}

\usepackage[acronym]{glossaries}
\usepackage{cleveref}
\usepackage{subcaption}
\usepackage{tcolorbox}
\usepackage{enumitem}

\usepackage{tikz}
\usepackage{pgfplots}
\usetikzlibrary{shapes.geometric, math, positioning}
\definecolor{ACMYellow}{RGB}{255, 214, 0}
\definecolor{ACMOrange}{RGB}{252, 146, 0}
\definecolor{ACMRed}{RGB}{253, 27, 20}
\definecolor{ACMLightBlue}{RGB}{131, 206, 226}
\definecolor{ACMGreen}{RGB}{166, 188, 9}
\definecolor{ACMPurple}{RGB}{101, 1, 107}
\definecolor{ACMDarkBlue}{RGB}{9, 53, 122}

\newacronym{icse}{ICSE}{International Conference on Software Engineering}
\newacronym{ase}{ASE}{Automated Software Engineering}
\newacronym{llm}{LLM}{Large Language Model}
\newacronym{llms}{LLMs}{Large Language Models}
\newacronym{se}{SE}{Software Engineering}

\newcommand{\RQOne}{To what extend are \acrshort{llm}-centric \acrshort{se} studies reproducible?}
\newcommand{\RQTwo}{Which factors impede the reproduction of \acrshort{llm}-centric empirical \acrshort{se} studies?}
\newcommand{\RQThree}{How well do the ACM artefact badges reflect the state of reproducibility of \acrshort{llm}-centric \acrshort{se} studies?}

\tcbset{colback=yellow!5!white,colframe=ACMYellow, coltitle=black}

\AtBeginDocument{%
  }

\copyrightyear{2026}
\acmYear{2026}
\setcopyright{cc}
\setcctype{by}
\acmConference[ICSE '26]{2026 IEEE/ACM 48th International Conference on Software Engineering}{April 12--18, 2026}{Rio de Janeiro, Brazil}
\acmBooktitle{2026 IEEE/ACM 48th International Conference on Software Engineering (ICSE '26), April 12--18, 2026, Rio de Janeiro, Brazil}
\acmPrice{}
\acmDOI{10.1145/3744916.3773207}
\acmISBN{979-8-4007-2025-3/2026/04}

\RequirePackage[
  datamodel=acmdatamodel,
  style=acmnumeric,
  sorting=none
  ]{biblatex}

\begin{document}

\title{Reflections on the Reproducibility of Commercial LLM Performance in Empirical Software Engineering Studies}

\author{Florian Angermeir}
\email{angermeir@fortiss.org}
\orcid{0000-0001-7903-8236}
\affiliation{%
  \institution{fortiss and Blekinge Institute of Technology}
  \city{Munich and Karlskrona}
  \country{Germany and Sweden}
}

\author{Maximilian Amougou}
\email{amougou@fortiss.org}
\orcid{0009-0003-3831-1007}
\affiliation{%
  \institution{fortiss}
  \city{Munich}
  \country{Germany}
}

\author{Mark Kreitz}
\email{mark.kreitz@unibw.de}
\orcid{0009-0001-8282-8930}
\affiliation{%
  \institution{University of the Bundeswehr Munich}
  \city{Munich}
  \country{Germany}}

\author{Andreas Bauer}
\email{andreas.bauer@bth.se}
\orcid{0000-0002-2916-4020}
\affiliation{%
  \institution{Blekinge Institute of Technology}
  \city{Karlskrona}
  \country{Sweden}}

\author{Matthias Linhuber}
\email{matthias.linhuber@tum.de}
\orcid{0000-0003-3640-5424}
\affiliation{%
 \institution{Technical University of Munich}
 \city{Munich}
 \country{Germany}}

\author{Davide Fucci}
\email{davide.fucci@bth.se}
\orcid{0000-0002-0679-4361}
\affiliation{%
  \institution{Blekinge Institute of Technology}
  \city{Karlskrona}
  \country{Sweden}}

\author{Fabiola Moyón C.}
\email{fabiola.moyon@siemens.com}
\orcid{0000-0003-0535-1371}
\affiliation{%
  \institution{Siemens AG}
  \city{Munich}
  \country{Germany}}

\author{Daniel Mendez}
\email{daniel.mendez@bth.se}
\orcid{0000-0003-0619-6027}
\affiliation{%
  \institution{Blekinge Institute of Technology and fortiss}
  \city{Karlskrona and Munich}
  \country{Sweden and Germany}}

\author{Tony Gorschek}
\email{tony.gorschek@bth.se}
\orcid{0000-0002-3646-235X}
\affiliation{%
  \institution{Blekinge Institute of Technology and fortiss}
  \city{Karlskrona and Munich}
  \country{Sweden and Germany}}

\renewcommand{\shortauthors}{Angermeir et al.}

\begin{abstract}
%%%%%%%%%%%Rationale
\acrlong{llm}s have gained remarkable interest in industry and academia. The increasing interest in \acrshort{llm}s in academia is also reflected in the number of publications on this topic over the last years. For instance, alone 78 of the around 425 publications at \acrshort{icse} 2024 performed experiments with \acrshort{llm}s.
%%%%%%%%%%%Problem
Conducting empirical studies with \acrshort{llms} remains challenging and raises questions on how to achieve reproducible results, for both researchers and practitioners. One important step towards excelling in empirical research on \acrshort{llm} and their application is to first understand to what extent current research results are eventually reproducible and what factors may impede reproducibility. This investigation is within the scope of our work.

%%%%%%%%%%%Contribution
We contribute an analysis of the reproducibility of \acrshort{llm}-centric studies, provide insights into the factors impeding reproducibility, and discuss suggestions on how to improve the current state.
%%%%%%%%%%%Methods
In particular, we studied the 85 articles describing \acrshort{llm}-centric studies, published at \acrshort{icse} 2024 and \acrshort{ase} 2024. Of the 85 articles, 18 provided research artefacts and used OpenAI models. We attempted to replicate those 18 studies.
%%%%%%%%%%%Conclusion
Of the 18 studies, only five were sufficiently complete and executable. For none of the five studies, we were able to fully reproduce the results. Two studies seemed to be partially reproducible, and three studies did not seem to be reproducible.
%%%%%%%%%%%Implications
Our results highlight not only the need for stricter research artefact evaluations but also for more robust study designs to ensure the reproducible value of future publications.
\end{abstract}




\maketitle

\section{Introduction} \label{sec:introduction}

\acrfull{llms} such as GPT, Gemini, and LLaMA have shown impressive capabilities for various software engineering tasks~\cite{zheng2024towards, 10.1145/3649506, 10440574}. Reflecting this trend, the \acrfull{se} research community has increasingly explored the use of \acrshort{llms} as both a subject of research and as tools for research~\cite{10.1145/3695988,10449667}. This trend is also evident in significant funding initiatives of more than 1\$ billion across the European Union\footnote{https://digital-strategy.ec.europa.eu/en/policies/genai4eu}, United States\footnote{https://www.nsf.gov/news/nsf-partners-kick-nairr-pilot-program}, and China, leading to research increasingly centered around, or enabled by \acrshort{llms}. In this work, when we refer to \acrshort{llms}, we mean autoregressive models (e.g., GPT, LLaMA, Mistral), as opposed to language models like BERT or T5, which are pretrained for different purposes.

A key characteristic of \acrshort{llms} is their non-deterministic nature, which manifests in different outputs for identical repeated queries to the same model. This non-determinism in combination with the lack of profound understanding of the factors influencing output performance, along with the rapidly evolving models, raises questions about the reproducibility of \acrshort{llm}-centric empirical \acrshort{se} studies. Several recent studies highlighted the difficulties of reproducing findings of \acrshort{llm}-centric empirical studies, in \acrshort{se} \cite{10.1145/3639476.3639764, vaugrante2024looming, 10.1145/3673791.3698432}, but also in other disciplines such as political science \cite{barrie2024replication}, and medicine \cite{krishna2024evaluation}. Some empirical studies explicitly acknowledge the unlikeliness of reproduction of their findings: ''Please note that due to the nondeterministic nature of LLMs, it is unlikely that exact replication of our results will be possible. However, similar results should be obtainable.''~\cite{10.1145/3691620.3695512}.
At this point, one could argue that already existing deep learning methods face the same reproduction issues given their inherent non-determinism. We agree that existing research addressed reproducibility in deep learning---e.g., \cite{10.1145/3477535}. However, the reproducibility of \acrshort{llms}-centric studies faces additional challenges such as predominant usage of proprietary models, frequent model updates without transparent changes, higher degree of configurability, and strong dependence on prompt engineering.

To this day, substantial resources have been invested into \acrshort{llm}-centric research--not only financially, but also in terms of environmental impact and academic attention. If those research results can not be reliably reproduced, these studies risk becoming time-bound artefacts providing a snapshot of the technology capabilities rather than building blocks of cumulative scientific knowledge. This concern is exacerbated by the historical lack of replication culture in \acrshort{se}, despite improvements in recent years~\cite{8894007, 10.1007/s10664-012-9227-7}.

To tackle this challenge in the long term, it is important to first understand the state of reproducibility of \acrshort{llm}-centric empirical \acrlong{se} studies and indicators towards enhanced or degraded reproducibility. This is in the scope of our study presented in this paper. In particular, we aim to answer the following research questions: 

\begin{enumerate}
    \item[\textbf{RQ1}] \RQOne
    \item[\textbf{RQ2}] \RQTwo
    \item[\textbf{RQ3}] \RQThree
\end{enumerate}

As a basis for our investigations, we analyse this at the example of publications presented at the \acrfull{icse} and \acrfull{ase}, both in their 2024 edition, considering the diversity of topics in their programs and research methods applied, but also due to the high ambition with respect to rigour, relevance, and validity. It is also important for us to note that our ambition is not to criticise individual contributions, let alone individual researchers. We believe that their work was guided by what may generally be considered to be of high scientific standard and, thus, allows us to critically reflect upon the overall state of reproducibility of \acrshort{llm}-centric \acrshort{se} research in the community and how to improve it.

\section{Background} \label{sec:background}
In this section, we define the relevant terms, discuss \acrshort{llms} inherent non-determinism and implications for reproducibility, give a brief introduction to Bayesian bootstrapping, the ACM badges, and our online material enabling reproducibility.

\subsection{Definitions}
We adopt the definitions of the ACM \cite{acm2020artifactreview}:
\begin{description}
    \item[Reproducibility:] ''\textit{The measurement can be obtained with stated precision by a different team using the same measurement procedure, the same measuring system, under the same operating conditions, in the same or a different location on multiple trials. For computational experiments, this means that an independent group can obtain the same result using the author’s own artifacts.}''
    \item[Replicability:] ''\textit{The measurement can be obtained with stated precision by a different team, a different measuring system, in a different location on multiple trials. For computational experiments, this means that an independent group can obtain the same result using artifacts which they develop completely independently.}''
\end{description}
In our work, as we aim to repeat studies that entail commercial \acrshort{llms}, several variables do not resonate with the ACM's definition of reproducibility, but also do not qualify for the definition of replicability, such as changes to \acrshort{llms}, and imperfect replication packages. However, the two terms have not been uniformly defined in the research communities, even contradicting each other \cite{national2019reproducibility, DEMAGALHAES201576}, while some authors argue that no distinction should be made between them \cite{gundersen2021fundamental, goodman2016does}. Given the grey area we interact in and our motivation to validate the results, we will---unless otherwise specified---use the terms interchangeably to refer to our act of repeating the studies and analysing the obtained results.

\subsection{Non-Determinism \& Reproducibility}
Assuming that reproduction serves as a means to uncover the true effects of a subject being studied, a high confirmatory reproduction rate of research findings is desirable. However, the confirmatory reproduction rate of research findings has historically been low in software engineering~\cite{10.1007/s10664-012-9227-7}, but also in more mature fields such as psychology~\cite{open2015estimating, schimmack2020meta}. 
In the context of \acrshort{llms}, various studies have acknowledged the importance and issues with replicability (e.g.,~\cite{lissack2024bridging, 10.1145/3639476.3639764}) even to the point of acknowledging potential non-replicability due to non-determinism, such as \cite{10.1145/3691620.3695512}.

%2. Non-Determinism in \acrshort{llms}
Given that non-determinism strongly contributes to the power of \acrlong{llm}s, evaluating the performance of an \acrshort{llm} in the light of non-determinism plays a vital role. To this day, we know some of the variables that influence non-determinism, such as temperature~\cite{song2024good}, prompting style~\cite{vaugrante2024looming}, or hardware~\cite{blackwell2024towards}. Despite partly contradictory investigations such as~\cite{11025952} and \cite{10.1145/3697010}, it remains unclear to what extent each factor influences non-determinism~\cite{song2024good}. This inherent non-determinism of \acrshort{llms} impacts the robustness of its performance~\cite{atil2024llm, klishevich2025measuring, 10.1145/3697010}, hence it impacts their reproducibility. 

To estimate the non-determinism of \acrshort{llms}, recent research focuses on quantifying the uncertainty associated with \acrshort{llms} when performing tasks.
To do so, research distinguishes between epistemic (lack of knowledge) and aleatoric (multiple solutions possible) uncertainties. There have been various approaches proposed to estimate uncertainties~\cite{huang2024survey}, such as \cite{yadkori2024believe} or \cite{ma2025estimating}. However, the inherent challenges of these approaches render their application infeasible for replication purposes. First, many approaches are bound to one perspective, like quantifying hallucinations.
Second, these approaches may involve modifying the original task (e.g., through iterative prompting like in \cite{yadkori2024believe}).
Finally, the uncertainty varies across different models~\cite{ma2025estimating} and tasks.
It appears to be unrealistic to develop a comprehensive function for estimating non-determinism.
This renders statistical power estimations---such as determining how many times an experiment with an \acrshort{llm} must be run to achieve a reliable, statistically significant result---only possible in small and highly controlled setups.
This limitation poses a significant threat to the validity of reproduction studies that seek comparability across multiple studies.

In addition to intrinsic factors of non-determinism, extrinsic time-related factors have a critical impact on \acrshort{llm} performance, in at least two ways.
First, \acrshort{llms} evolve rapidly over time through regular updates, leading to different model versions, such as those reflected in OpenAI's minor versioning system (e.g., \textit{gpt-4o-2024-05-13} is a minor version of the \textit{gpt-4o-model}).
This change in model versions can impact the consistency of \acrshort{llms} performance, depending on the individual task and context. 
Academic literature \cite{10.1145/3644815.3644950, chen2023chatgptsbehaviorchangingtime}, as well as literature by model providers \cite{openai2025functioncalling}, acknowledge this time-related factor.
For example, OpenAI tries to mitigate negative performance impacts through their own and community-sponsored benchmarks.
However, those benchmarks cannot cover the broad variety of \acrshort{llm} tasks defined in the research community.
Second, for time-dependent tasks in which information gets outdated fast, older models may provide less relevant information, resulting in decreased performance. 
It remains unclear how the general performance of a fixed version of a model evolves over time, particularly for \acrshort{llms} of commercial providers. For instance, OpenAI claims that a ''\textit{pinned model is stable, meaning that we won’t make changes that impact the outputs}''~\cite{openai2025functioncalling}.
However, it has yet to be evidently shown to what extent this promise holds in practice.

\subsection{Baysian Bootstrap}\label{subsec:baysian}
The Bayesian bootstrap~\cite{rubin1981bayesian} is a method to generate insights into data without knowing its underlying distribution. It does so by simulating multiple random draws from a potential distribution over the existing data. It differs insofar from non-parametric bootstraping as it does not resample the data, but rather weights the individual data. For a general introduction and visualisation, consider the original publication by \textcite{rubin1981bayesian} or this blogpost by \textcite{baath2015bootstrapbayesian}. In our work, we focus on the analysis of the confidence intervals over the data instead of on a specific statistic, as is done traditionally in the Bayesian bootstrap. For that purpose, we follow the procedure described in \cite{musio2006bayesian}. We run the procedure 10.000 times and calculate the mean for percentiles over all estimations to obtain the final percentile value. We do this analysis for the following percentiles: 2.5th, 5th, 10th, 25th, 50th, 75th, 90th, 95th, 97.5th. This allows us to set the originally-reported values into perspective of the values obtained through reproduction, allowing us to make statements within a 95\% interval, within which we consider the reproduction to be likely. Outside this 95\% interval, we perceive the reproduction as unlikely. 

\subsection{Reproduction Framework}
As our work faces the risk of getting outdated fast, we aim to provide our work in a reproducible fashion, potentially for future meta-analysis. For that purpose, we developed a simple reproduction framework that enables direct execution of the analysed studies.
The framework has been built by five of the authors. It is based on containerisation to ensure platform independence and version stability---to the extent possible. Each reproduced study is integrated in a way that the research artefacts are downloaded on demand, and, if necessary, modified transparently by the reproduction framework. This has two advantages. First, we do not infringe intellectual property (e.g., if authors did not publish the artefacts with a license), and second, the changes are transparent, hence they can be verified---or challenged. The reproduction framework is openly available here \cite{replikit}.

\subsection{ACM Badges}
ACM offers three major types of badges~\cite{acm2020artifactreview}: The \textit{artifacts available}, \textit{artifacts evaluated}, and \textit{results validated} badges.
The available badge signifies that the associated artefacts are archived permanently. The evaluated badges are for those papers that provide artefacts to the reviewers for critical review of the artefacts documentation, consistency, completeness, and exercisability.
Finally, the results validated badges are for those papers, whose results have been successfully confirmed in a separate study.
The review process for each of those badges is not prescribed by ACM, and its interpretation is to some extent left open to those chairing individual venues. For example \acrshort{icse} 2021 deviated from ACM's requirements\footnote{\url{https://conf.researchr.org/track/icse-2021/icse-2021-Artifact-Evaluation\#Call-For-Artifact-Submissions}}.

\section{Methodology} \label{sec:methodology}
\begin{figure}[thb]
    \centering
    \begin{tikzpicture}[font=\small]

\tikzmath{
\ySpace= 0.7;
}

\tikzset{%
	a/.style={thick},
	box/.style={
		rectangle,
		draw,
		rounded corners,
		minimum height=2em,
		text width=8em,
		text centered,
	},
	txt/.style={font=\small\itshape},
    % Left = green, Right = orange
    txtL/.style={
        font=\small\itshape,
        text=green!60!black
    },
    txtR/.style={
        font=\small\itshape,
        text=orange!80!black
    },
	database/.style={
      cylinder,
      shape border rotate=90,
      aspect=0.25,
      draw
    }
}

% ICSE
\node[database] (dbI) at (0,0) {ICSE};
\node[box, below=\ySpace of dbI] (icI) {Search};
\node[box, below=\ySpace*3 of icI] (icM) {Manual Exclusion};

\draw[a, ->] (dbI) -- (icI) node[txtL, left, pos=0.5] {425 articles};
\draw[a, ->] (icI) -- (icM) node[txtL, left, pos=0.5] {109 articles};

% ASE
\node[database] (dbA) at (4,0) {ASE};
\node[box, below=\ySpace of dbA] (icA) {Search};
\node[box, below=\ySpace of icA] (icAbadge) {Exclusion: no artifact badge};
\node[box, below=\ySpace*3 of icA] (icAM) {Manual Exclusion};

\draw[a, ->] (dbA) -- (icA) node[txtR, right, pos=0.5] {314 articles};
\draw[a, ->] (icA) -- (icAbadge) node[txtR, right, pos=0.5] {108 articles};
\draw[a, ->] (icAbadge) -- (icAM) node[txtR, right, pos=0.5] {7 articles};

% Data extraction
\node[box, below=\ySpace*5 of icI] (ext) {Extract data};
\node[box, below=\ySpace of ext] (exOpenAI) {Inclusion/Exclusion: OpenAI usage};
\node[box, below=\ySpace of exOpenAI] (exRep){Seemingly fit for reproduction};
\node[box, below=\ySpace of exRep] (reproduced){Reproduction by five researcher};
\node[below=\ySpace of reproduced] (end) {\textbf{5 articles fit for reproduction}};

\draw[a, dotted, ->] (exRep.east) -- ++(4.5,0) |- (dbA.east) node[txt, above, pos=0.75] {motivates};

\draw[a, ->] (icM) -- (ext) node[txtL, left, pos=0.5] {78 articles};
\draw[a, ->] (icAM) |- (ext) node[txtR, right,pos=0.2] {7 articles};

\draw[a, ->] (ext) -- (exOpenAI)
    node[txtL, left,  pos=0.35] {78 articles}
    node[txtR, right, pos=0.70] {7 articles};

\draw[a, ->] (exOpenAI) -- (exRep)
    node[txtL, left,  pos=0.35] {63 articles}
    node[txtR, right, pos=0.70] {6 articles};

\draw[a, ->] (exRep) -- (reproduced)
    node[txtL, left,  pos=0.35] {13 articles}
    node[txtR, right, pos=0.70] {5 articles};

\draw[a, ->] (reproduced) -- (end);
\end{tikzpicture}
    \caption{Summary of Research Approach}
    \label{fig:research-approach}
\end{figure}
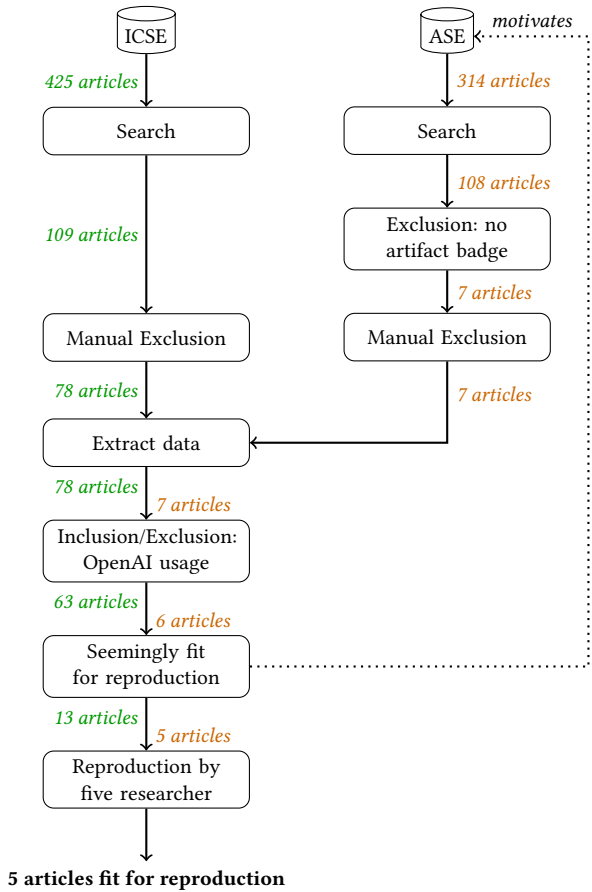

We analysed, through a critical review~\cite{10.1145/3540250.3560877} and reproduction analysis, \acrshort{llm}-centric empirical software engineering studies from the highly influential software engineering conferences: \acrlong{icse} 2024 and \acrlong{ase} 2024. We reproduced studies where feasible and provided a reproduction package at \cite{online-material}.
\Cref{fig:research-approach} provides a visual summary of our research approach.

\subsection{Literature Review}

We conducted a literature review in the form of a critical review as described by \textcite{10.1145/3540250.3560877} to identify relevant articles containing LLM replication packages.

\paragraph{Article sources}
As sources for articles, we choose the proceedings of \acrshort{icse} 2024~\cite{10.1145/3597503}, as the representative conference for the software engineering community. Given the resource intensity of this research, we decided against conferences with focus on specific aspects of software engineering such as MSR or ICSME.

We also did not consider the proceedings of previous years, as they did not contain relevant studies at scale and due to concerns about the timeliness of those results.
For the same timeliness reason we did not opt for journals, which have a longer review cycle. 
At the time this study was conducted, the ICSE proceedings for 2025 had not been published, hence newer studies were not available for analysis.

Of ICSE 2024, we considered the following tracks for this study: research, companion, nier, seis, and seip.
Due to the low number of available and executable research artefacts, we also included the proceedings of \acrshort{ase} 2024~\cite{10.1145/3691620}, as \acrshort{ase} is the top venue for automated software engineering, which aligns well with the notion of LLMs as automation and tooling.

\paragraph{Search Query \& Filtering}
To identify relevant articles, we searched through the 425 ICSE articles directly via the proceedings PDFs using keywords in the title, abstract, and author keywords. 
For the keywords, we iteratively defined a set of keywords based on our experience of identifying relevant literature during our earlier contributions to the LLM guidelines \cite{baltes2025guidelinesempiricalstudiessoftware}.
This left us with a final set of the following keywords:
\textit{genai, llm, large language model, prompt, chain-of-thought, chain of thought, zero-shot, zero shot, few-shot, few shot, llama, claude, mistral, gemini, openai, gpt}. The automatic search resulted in 109 articles.
We reviewed a sample of the automatically excluded articles to ensure no relevant literature was missed.
To cover as much empirical research as possible, the initial exclusion focused on (1) non-empirical research, and (2) research with non-autoregressive \acrshort{llm}s (e.g., BERT). Two authors manually reviewed the 109 articles and identified 31 articles that were out of our scope, resulting in 78 relevant articles.

Later, we complemented the low number of suitable articles from \acrshort{icse} with those from \acrshort{ase} in an iterative process.
In this iterative process, we did not aim for completeness, but instead focused on selected articles with an ACM artefact badge, to strengthen our critical review~\cite{10.1145/3540250.3560877} while adding the angle of automation, especially to strengthen the results of RQ3. Here we operated under the assumption, that ACM badges would guarantee a high artefact quality, hence signal a higher readiness for reproduction. Whether this holds true is to be seen in \Cref{sec:results}. This process resulted in seven relevant ASE articles.

\paragraph{Data Extraction}
We iteratively extracted all relevant metadata of articles, as well as the published research artefacts, to select suitable articles and collected them into a spreadsheet. 
This includes the following information:
    (a) conducted LLM experiment,
    (b) contains ACM artefact badge,
    (c) LLM name,
    (d) availability of research artefact/data,
    (e) performs a comparison between LLMs,
    (f) defines temperature,
    (g) report top-p, top-k, 
    (h) report prompts,
    (i) handles context window limitations, and
    (j) detailed model version. 
    
\paragraph{Exclusion of Articles}
As the extraction revealed, over 70\% of the articles utilised \acrshort{llms} provided by OpenAI as the primary model or as a benchmark model.
Complemented with the motivation to provide a unified setup that facilitates reproduction, validation, and extension of our work, we excluded all articles that did not use OpenAI's models.
Although this limits the generalisability of our findings to a single model provider, it aligns with the focus of our research—commercial \acrshort{llm}s.
Additionally, OpenAI models were used by a significant portion of the analysed articles.

Next, we excluded all articles that did not provide a research artefact or data. 
Over two-thirds of the articles either lacked a reference to a research artefact at all or were incomplete so that the retrieval of artefacts was not possible.
This encouraged us to perform a second iteration, with articles published at \acrshort{ase}, where we additionally excluded articles without an ACM artefact badge to counter the issue of missing or incomplete artefacts resulting in a total of 85 articles, of which 69 used OpenAI models. From those 69 articles, 18 seemed to provide complete and executable research artefacts and data at first inspection. We call those fit for reproduction.

\subsection{Analysis}
We perform the analysis from three different perspectives, each laid out in the following.
\paragraph{RQ1}
For RQ1, we used the 18 articles that seemed fit for reproduction. 
To that end, a team of five researchers developed a framework to systematically execute replication code for analysing and reproducing research artefacts.
To ensure platform independence, we provide a stable environment by containerising all study experiments.
The containerised version of the experiments makes its dependencies explicit, where missing dependencies were supplemented with suitable versions.
Each of the five authors independently analysed the repeatability of at least two of the 18 studies adhering to the following process:
\begin{enumerate}
    \item Access the research artefact and verify if the code and data are complete for reuse according to the ACM definition of reusability~\cite{acm2020artifactreview}, this rendered 13 articles ill-suited for reproduction.
    \item Execute the research artefact as it is, document challenges, and address any hurdles, provided that the modifications do not impact the experiment's functionality. We will discuss these alterations in the Threats to Validity section.
    \item Create automation tasks for the research artefacts within the framework.
    \item Multiple executions of replication artefacts within the framework.
    \item Document observations.
\end{enumerate}

To minimise researcher bias and ensure functional correctness, portability, and understandability, each study's integration and modification were reviewed by another co-author before being integrated into the framework. This code review was facilitated by initial meetings of all authors involved in the reproduction process. Those meetings led to a mutual understanding regarding the process and the review criteria among the authors. The reviews focused on the correctness of the automation, containerisation, portability to other environments, and the clarity of interventions, comments, results and documentation. 
We documented the changes transparently in our supplementary online material~\cite{online-material}.

As we discussed in \Cref{sec:background}, estimating the required number of artefact runs per study to ensure statistical significance is not feasible.
However, we can attempt to repeat each experiment run until its prediction interval is below a certain threshold~\cite{blackwell2024towards}.
While this approach is suitable for individual studies, it can become computationally expensive in the context of multiple studies.
This is due to the high costs associated with each experimental run, which can reach up to \$600, as noted in~\cite{10.1145/3691620.3695022}.
Hence, we chose to adopt a more practical approach. Each study was conducted up to 30 times or until we reached a cost of \$500.
Although it may not be statistically significant, we consider 30 repetitions to be a sufficiently large sample size to make a practical statement about the likelihood of successfully reproducing the research results.
The cost cap of \$500 was dictated by the budget for the research, taking into consideration the number of executable studies. The costs were monitored through the OpenAI API dashboard. This cost cap impacted one study (\cite{10.1145/3597503.3639226}). We report the reproduction costs for each study in \Cref{sec:results}.
We have not contacted the authors of the studies for replication to assist in recovering artefacts and replicating results.
Help by the original authors could introduce potential bias~\cite{DOSSANTOS2022106870} and contradict our aim of independently replicating results.

After reproduction, we analyse the results from two perspectives: descriptively and using Bayesian bootstrapping.
As descriptive analysis, we report the difference between \acrshort{llm} performance metrics observed by us and reported in the articles.
Furthermore, we provide a full list and a graphical representation of all reproduced metrics in our online material~\cite{online-material}.
To account for the inherent uncertainty in the \acrshort{llm}'s performance, we employ a version of Bayesian bootstrapping focused on confidence intervals~\cite{musio2006bayesian} as described in \Cref{subsec:baysian}.
We apply Bayesian bootstrapping on the observed metrics to understand whether the originally reported metrics are likely in light of the observed metrics. 

\paragraph{RQ2} For RQ2, we took two perspectives. A meta-analytical one, and an experimental one. For the meta-analytic perspective, we analysed the data extracted from all 85 articles to understand the general reasons for non-reproducibility. In the experimental one, we took the 18 articles that provided a research artefact and aimed to understand the factors that directly impact reproducibility during reproduction.

For the meta-analytic perspective, we first analysed the availability and completeness of research artefacts and data as the prerequisites for reproducibility. For a deeper analysis, we oriented ourselves on the guidelines 5.2, 5.3, and 5.4 of the LLM guidelines for software engineering~\cite{baltes2025guidelinesempiricalstudiessoftware}. Specifically, we investigated the reporting of the model versions, the temperature as well as top-p/top-k parameters as representatives for the model configuration (guideline 5.2). As representative for guideline 5.3, we investigated the description of context window considerations as part of a mindful architecture to ensure robust reuse. Finally, we checked whether prompts were reported in the article, or where not feasible, described in prompt templates (guideline 5.4). Those characteristics do not cover the LLM guidelines in their entirety, but provide first insights into the current state of practice, especially under the hypothesis, that adherence to those guidelines will improve reproducibility. Two authors collaboratively extracted those characteristics.

We approached the experimental perspective from two angles. The first angle investigated the existence of the before described characteristics in the research artefacts of the 18 articles. The second angle was fed by the issues faced during the reproduction attempts in RQ1. Each author independently documented the issues faced during the reproduction attempts. Afterwards, those issues were discussed among the authors and aggregated for reporting.

\paragraph{RQ3} For RQ3, we examined the association between ACM artefact badges and the reproducibility attempts of the 18 articles. For that purpose, two authors coded all articles based on their assigned badges: \textit{Artifacts Available v1.1}, \textit{Artifacts Evaluated - Functional v1.1}, and \textit{Artifacts Evaluated - Reusable v1.1}. Informed by the outcomes of the reproduction efforts from RQ1, and the analysis of factors impeding reproduction, we drew conclusions on whether the badge types reliably signaled reproducibility.

The resulting data for all analyses are available in the online material \cite{online-material}.

\section{Results} \label{sec:results}
\begin{figure}[htb!]
    \centering
    \resizebox{\columnwidth}{!}{%
\begin{tikzpicture}

\begin{axis}  
[  
    xbar,  
    enlargelimits=0.1,  
    xlabel={\# Occurences}, 
    symbolic y coords={ gpt-4-0613, gpt-2, code-davinci-002, codegen, starcoder, chatgpt, text-davinci-003, gpt-3.5, gpt-3.5-turbo, gpt-4,}, 
    ytick=data,  
     nodes near coords,
     nodes near coords align={horizontal},  
    ]  
\addplot [fill=ACMLightBlue] coordinates {
	(23,gpt-4)
	(17,gpt-3.5-turbo)
	(14,gpt-3.5)
	(8,text-davinci-003)
	(7,chatgpt)
	(6,starcoder)
	(5,codegen)
	(5,code-davinci-002)
	(3,gpt-2)
	(3,gpt-4-0613)
	};  
\end{axis}  
\end{tikzpicture}  
}
    \caption{Top 10 \acrshort{llms} Employed in the Analyzed 85 Articles.}
    \label{fig:top15models}
\end{figure}
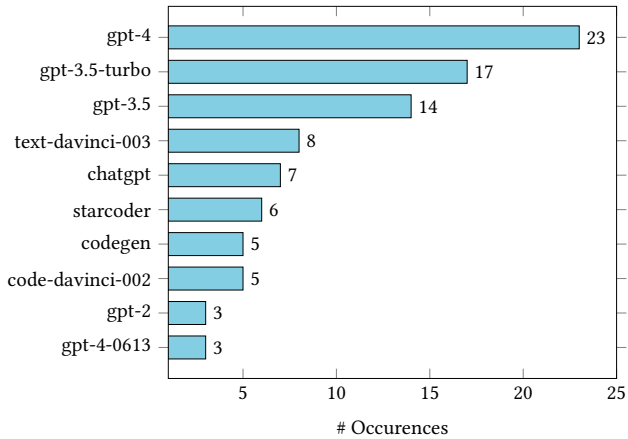

For \acrshort{icse} 2024, the search yielded 109 articles. Of those 109 articles, 78 performed experiments with LLMs. At \acrshort{ase} 2024, 108 articles performed experiments with LLMs. Of those, seven articles were awarded an ACM artefact badge. Hence, we analysed a total of 85 articles. Of those 85 articles, 69 employed OpenAI services as their base model or for comparison. In total, 43 of 85 articles compared multiple \acrshort{llms}. \Cref{fig:top15models} depicts the top 10 models used in the 85 papers, highlighting again the dominance of OpenAI services in academia as per 2024.

\subsection{RQ1: \RQOne}
Out of 69 studies using OpenAI services, only five studies were fit for reproduction. Below, we briefly introduce each, identify reproducibility-impacting variables, and summarise our findings. Details on the factors rendering the remaining 64 studies unfit for reproduction can be found under RQ2.

\paragraph{Reproduction of \cite{10.1145/3597503.3639226}}
The study by \textcite{10.1145/3597503.3639226} investigated the capabilities of \acrshort{llms} to translate code from one programming language to other programming languages. They tested seven \acrshort{llms} for their performance of correct translation over five datasets. Given the size of the experiment---and the associated costs---we ran a subset of those experiments: We repeated their experiments of translating from Python and Go to the other programming languages over the dataset CodeNet (see Table 2 in the original study). Given that each experiment repetition costs approximately \$30, we decided to run 15 repetitions, amounting to a total of approximately \$450. The results and the generated evidence can be observed in our online material \cite{online-material}. \\
\textbf{Variables impacting reproduction:} The paper did not come with a pre-built run-time for all five programming languages. Using the versions mentioned in their paper, we containerised a run-time offered for review in our online material \cite{online-material}. Additionally, no minor version for the employed model (GPT-4) was specified, impeding a direct reproduction. We used gpt-4-0613. \\
\textbf{Results:} For the translation from Python to other programming languages, Pan et al. reported an average successful translation ratio of 79.9\%. Given the average successful translation ratios we obtained in our 15 runs, we calculated a 95\% interval range from 45.6\% to 55.3\%. For the translation from Go to other programming languages, we arrived at a 95\% interval ranging from 50.1\% to 54.5\% with an originally reported value of 85.5\% of average successful translations. The maximum average successful translation ratio we saw for Python and Go was 56\% and 55\%. \\
\textbf{Summary:} For the experiments we repeated, we could not only not reproduce the results, but are far off with our results as becomes apparent from the \cite{10.1145/3597503.3639226}-section of \Cref{tab:all_ranges}.

\paragraph{Reproduction of \cite{10.1145/3597503.3639183}}
The study by \textcite{10.1145/3597503.3639183} evaluates the effect of three different automatic semantic augmentations for prompts on code summarisation tasks with six different programming languages.
As a baseline, BM25 prompting is used, which is a few-shot prompting approach leveraged by exemplars that are "semantically close" to the target \cite{10.1109/ICSE48619.2023.00205}.
All augmentations combined on top of BM25 form the presented ASAP method. \\
\textbf{Variables impacting reproduction:} Most of the study was performed on the deprecated \textit{code-davinci-002} model. Thus, this part, which, e.g., included the experiments on the individual impact of each augmentation, cannot be reproduced. However, a comparison between different LLMs is made, evaluating the effectiveness of the ASAP method for summarisations of Java, Python, and PHP code. Of the three models used, only \textit{gpt-3.5-turbo} is still available. The authors did not state a specific model version, neither in the paper nor in the research artefact. We used \textit{gpt-3.5-turbo-0125} for reproduction. Furthermore, no explicit instructions on how to run the experiments with the ASAP method were given. We assumed to run it by enabling the feature flag for all augmentations. Lastly, versions for the dependencies in the code were not stated, causing issues with deprecated functions. These could be solved by installing older versions of the affected libraries. \\
\textbf{Results:} We ran 30 repetitions of the experiment, which amounted to approximately \$30. \Cref{tab:all_ranges} shows that none of the results reported by \cite{10.1145/3597503.3639183} are within our 95\% intervals. Our data shows that the code summarisation of \textit{gpt-3.5-turbo} for the baseline task (BM25) significantly outperforms the reported results for Java and Python, while falling behind in PHP. 
The ASAP method still slightly outperforms the baseline in Java. However, the results have drastically changed for Python and PHP, where it is now underperforming. The strong changes indicate that the possibly newer version of \textit{gpt-3.5-turbo} we used requires different prompting than the one used in the original experiments. \\ 
\textbf{Summary:} None of the results that use available models could be reproduced within our experiments.

\paragraph{Reproduction of \cite{10.1145/3639476.3639757}}
The study of \textcite{10.1145/3639476.3639757} evaluated the capabilities of \acrshort{llms} to generate regular expressions from natural language prompts and assessed the functional correctness and security (i.e., resistance to ReDoS attacks) of generated regexes.
For each prompt, the authors tested regexes generated by four LLMs.
They evaluated the generated regexes using four metrics (exact match, DFA-equivalence@k, and pass@k, vulnerable@k).
The results and the generated evidence can be observed in our online material \cite{online-material}.\\
We repeated the experiment 30 times, amounting to approximately \$30. \\
\textbf{Variables impacting reproduction:}
Text-DaVinci-003 was already deprecated when we started our experiment. Although the paper specifies that the June 2023 version of GPT-3.5-Turbo was used, this was not indicated in the released code. Hence, we used the current model gpt-3.5-turbo-0125.
In addition, the artefact lacked the evaluation code. We could reconstruct the evaluation based on the provided documentation using the default configurations. We were not able to reconstruct the metric \textit{unparseable}. The changes are transparent in the online material \cite{online-material}. \\
\textbf{Results:}
Five of the originally reported values are likely to be reproducible (see \cite{10.1145/3639476.3639757} of \Cref{tab:all_ranges} for two examples) with the following minimum/maximum values: \textit{DFA-EQ@3 Refined} (11.55/15.49), \textit{vul@$1_{Rescue}$} (0.25/0.45), \textit{vul@$3_{Rescue}$} (0.83/1.41), \textit{vul@$10_{Rescue}$} (1.31 /2.36), \textit{vul@$10_{ReDoSHunter}$} (8.77/11.40). The small windows between minimum and maximum observed values speak for the stability of the results. The remaining 21 of the originally reported values were outside of the calculated 95\% interval. For those values that could not be reproduced, we saw at least the same trends, such as a performance increase between DFA-EQ@1 and DFA-EQ@10. \\
\textbf{Summary:} The majority of the results are unlikely to be reproducible.

\paragraph{Reproduction of \cite{10.1145/3639476.3639762}}
The study by \textcite{10.1145/3639476.3639762} investigated the effectiveness of large language models---specifically GPT-3.5 and GPT-4---in detecting vulnerabilities in C/C++ code for different prompts and configurations. They compared these models to a state-of-the-art baseline (CodeBERT) using \textit{in-context learning}. They performed their experiments twice to account for variability in model output.
We reproduced a subset of their configurations.  We ran their experiment 30 times for 8 different prompt configurations using GPT-3.5. Amounting to approximately \$36.\\
\textbf{Variables impacting reproduction:} Three configurations were out of scope for the reproduction due to unhandled context length limitations and lacking error handling (configuration A54, A4 k=3, A4 k=5). As GPT-4 was only applied to half of the test set due to cost constraints in the original paper, we omit its reproduction here. The artefacts had several general code issues, dependency versions were not specified, and prompt configurations were injected manually, which we automated. The model version was not specified. Thus, we employed gpt-3.5-turbo-0125. All changes, the results, and the generated evidence are transparently documented in our online material \cite{online-material}.\\
\textbf{Results:} Two-thirds of the original results seem to be reproducible. However, we saw vast differences in stability for the metrics. While some metrics were consistently in a 3\% range, others were found in a 10-30\% range. For example, the metric precision for the configuration \textit{P+A4 K=1} had a minimum observed value of 50\% and a maximum value of 82.35\%. In contrast to that, the metric accuracy for the same configuration had a minimum value of 50\% and a maximum value of 54.15\%. Those ranges are also represented in the \cite{10.1145/3639476.3639762}-section of \Cref{tab:all_ranges}. Given those varieties in the value ranges observed during the 30 experimental executions, the claim of reproducibility---even though in accordance with ACM's definition---has to be taken with a grain of salt here. \\
\textbf{Summary:} The majority of the results are likely to be reproducible.

\paragraph{Reproduction of \cite{10.1145/3691620.3694994}}
The research of \textcite{10.1145/3691620.3694994} proposes a \acrshort{llm}-based log parser. They tested their approach on two large-scale datasets against other state-of-the-art parsers. Among others, they measure the parser's performance over multiple metrics. For the reproduction, we focus on the LogHub-2k dataset experiments (Table 1 of the original article). The reason is that 1. the dataset was already in the provided research artefact, and 2. loading the other dataset would have required modifying the container as it lacked needed libraries. We repeated the experiments 30 times, amounting to approximately \$40. The results and the generated evidence can be observed in our online material \cite{online-material}. \\
\textbf{Variables impacting reproduction:} The research artefact does not generate one of the metrics (MLA), hence we could not reproduce this metric. \\
\textbf{Results:} 
In total, we reproduced two metrics (GA, ED) over 16 sub-datasets. As these are 32 combinations, we will here only give a summary of the reproducibility results. For 25 of the 32 combinations, the originally reported values were within the 95\% ranges we found in 30 repetitions. For two combinations (HPC-ED, Apache-ED, OpenStack-ED), the reported values were close (0.995, 0.999, 0.993) to the observed minimum/maximum values (0.996, 1.0, 0.992), and for five combinations (Windows-ED, Android-ED, MAC-GA, MAC-ED), the reported values are unlikely to be reproducible as they were below the values we observed (0.857 vs. 0.862, 0.955 vs. 0.966, 0.857 vs. 0.912, 0.881 vs. 0.897).\\
\textbf{Summary:} The majority of the original results are likely to be reproducible. Those original results that seem not reproducible were still close to the originally reported values as depicted in the \cite{10.1145/3691620.3694994}-section of \Cref{tab:all_ranges}.

\begin{table*}[htp]
    \centering
    \caption{Example Metrics from the Five Reproduced Studies. In Bold, the Percentiles Closest to the Originally Reported Values; in Red, the Original Values Falling Outside the 95\% Interval of Reproduced Values.}
    \begin{tabular}{@{}lp{3.9cm}cccccccc@{}}
    \toprule
    \textbf{Study} & \textbf{Configuration/Metric} & \textbf{Original Value} & \textbf{2.5th} & \textbf{10th} & \textbf{25th} & \textbf{50th} & \textbf{75th} & \textbf{90th} & \textbf{97.5th} \\
    \midrule
    \textcite{10.1145/3597503.3639226} & Successful Transl. from Python & \textcolor{red}{79.9} & 45.585 & 48.445 & 50.611 & 51.883 & 52.699 & 53.864 & \textbf{55.272} \\
    & Successful Transl. from Go & \textcolor{red}{85.5} & 50.118 & 50.388 & 50.812 & 51.652 & 52.838 & 53.702 & \textbf{54.520} \\
    \midrule
    \textcite{10.1145/3597503.3639183} & %Java BM25 & \textcolor{red}{16.68} & \textbf{18.540} & 18.705 & 18.840 & 19.019 & 19.222 & 19.395 & 19.608 \\
      Java ASAP & \textcolor{red}{16.96} & \textbf{18.954} & 19.016 & 19.128 & 19.239 & 19.302 & 19.437 & 19.627 \\
      %& Python BM25 & \textcolor{red}{15.01} & \textbf{15.746} & 15.775 & 15.841 & 15.954 & 16.154 & 16.286 & 16.382 \\
      & Python ASAP & \textcolor{red}{16.38} & 14.505 & 14.655 & 14.857 & 15.033 & 15.194 & 15.338 & \textbf{15.487} \\
      %& PHP BM25 & \textcolor{red}{18.48} & 17.376 & 17.482 & 17.599 & 17.729 & 17.867 & 18.001 & \textbf{18.061} \\
      & PHP ASAP & \textcolor{red}{19.99} & 16.615 & 16.667 & 16.712 & 16.817 & 16.906 & 16.964 & \textbf{17.012} \\
    \midrule
    \textcite{10.1145/3639476.3639757} & 
     vul@$1_{Rescue}$ & 0.28 & 0.250 & \textbf{0.262} & \textbf{0.289} & 0.306 & 0.347 & 0.393 & 0.428 \\
     & vul@$10_{Rescue}$ & 1.84 & 1.334 & 1.418 & 1.481 & \textbf{1.640} & \textbf{1.941} & 2.111 & 2.272 \\
     & DFA-EQ@1 Original &  \textcolor{red}{11.8} & 7.524 & 7.691 & 7.818 & 8.170 & 8.650 & 8.995 & \textbf{9.164} \\
     & vul@$3_{ReDoSHunter}$ Original &  \textcolor{red}{6.20} & 5.117 & 5.273 & 5.446 & 5.622 & 5.802 & 6.022 & \textbf{6.090} \\
    \midrule
    \textcite{10.1145/3639476.3639762} & P+A4 K=1 Precision & 75.2 & 53.10 & 61.42 & 65.59 & 70.38 & \textbf{74.69} & \textbf{78.23} & 81.49 \\
        & P+A4 K=1 Accuracy & 51.8 & 50.3 & 51.17 & \textbf{51.77} & \textbf{52.27} & 52.77 & 53.15 & 53.82 \\
        & P+A5 K=5 Recall & 47.2 & \textbf{46.73} & \textbf{47.47} & 49.17 & 50.97 & 52.60 & 52.60 & 56.50 \\
    \midrule
    \textcite{10.1145/3691620.3694994} & Proxifier GA & 1.000 & 1.0 & 1.0 & 1.0 & \textbf{1.0} & 1.0 & 1.0 & 1.0 \\
        & Hadoop ED & 0.953 & 0.952 & 0.953 & 0.953 & \textbf{0.953} & 0.953 & 0.953 & 0.953 \\
        & OpenStack GA & \textcolor{red}{0.968} & \textbf{0.971} & 0.985 & 0.989 & 0.99 & 0.99 & 0.99 & 0.99 \\
        & Mac ED & \textcolor{red}{0.881} & \textbf{0.896} & 0.898 & 0.899 & 0.900 & 0.901 & 0.902 & 0.902 \\
    \bottomrule
    \end{tabular}
    \label{tab:all_ranges}
\end{table*}

\begin{tcolorbox}[title=RQ1 Summary]
Of 69 studies, only five were fit for reproduction. None of those five studies could be completely reproduced. Two studies could be partially reproduced, and three studies could not be reproduced.
\end{tcolorbox}

\subsection{RQ2: \RQTwo} 
We look at the factors from two perspectives. First, from a meta-analytical perspective, and second, from an experimental perspective. For the meta-analysis, we look at all 85 articles. For the experimental one, we focus on the 18 articles that seemed fit for reproduction.

Of the 85 articles, 35 did not provide their code or data. For another 15 articles, those research artefacts were incomplete to the point of no recovery. Those two factors rendered 50 articles non-reproducible. Strictly following ACM's definitions, this does not imply non-replicability. If a paper describes the method and setup in detail enough, a replication might still be feasible. Of the 85 articles, 8 did not state the models used, describing them in a generic fashion as ''LLM'' or ''ChatGPT'', ''LLM-based''. Such works can, of course, not be replicated. Of the 85 articles, 29 reported the temperature configuration of their models, and only 14 reported on the configuration of other parameters, such as top-p and top-k. At the same time, only 22 articles described the context window size limitations and how they were handled. Finally, 50 articles reported on the prompts, either detailing the prompts or explaining the prompt templates, if the prompts were dynamic or too long to report entirely. \\
In the experimental investigation we evaluated those articles, that seemed fit for reproduction and that used OpenAI services. This resulted in 18 articles~\cite{10.1145/3597503.3623298, 10.1145/3639475.3640112, 10.1145/3639476.3639762, 10.1145/3639476.3639757, 10.1145/3639476.3639777, 10.1145/3597503.3639183, 10.1145/3597503.3639184, 10.1145/3597503.3608134, 10.1145/3597503.3639226, 10.1145/3597503.3639103, 10.1145/3597503.3608132, 10.1145/3597503.3623316, 10.1145/3597503.3639194, 10.1145/3691620.3695024, 10.1145/3691620.3695055, 10.1145/3691620.3695022, 10.1145/3691620.3694994, 10.1145/3691620.3694982}.
Of those articles, ten did not define the minor version of the \acrshort{llm} used in the research artefact. Three did not configure the temperature of the model. Ten did configure top-p and top-k. The context window was only handled by 5 of 18 research artefacts.

During reproduction, we ran into various issues for almost all of the 18 articles. Those issues are listed in \Cref{tab:failure-reasons}. The biggest issue we faced was non-obvious missing artefacts, such as individual scripts missing, and the importing of non-existent modules. The second most occurring issue was unspecified dependency versions. For some articles, this issue was fixable. For others, this resulted in dependency conflicts. General code issues were another reason for reproduction failure. Those were reflected in the usage of unassigned variables and hardcoded paths. Also, a major cause for reproduction failure was the deprecation of \acrshort{llms}. In at least four studies, we faced the issue of deprecated models, which, of course, we could not solve. Other issues were incomplete or missing documentation on how to reproduce the results, non-executable artefacts, lacking error handling---especially for long-running experiments---and deprecated external API points.

\begin{table}[ht]
\centering
\caption{Factors Impeding Reproduction of the 18 Articles That Seemed Fit for Reproduction}
\begin{tabular}{@{}p{0.6\columnwidth}l@{}}
\toprule
\textbf{Factor} & \textbf{Articles} \\ \midrule
Incomplete Artefacts & \cite{10.1145/3691620.3694982, 10.1145/3691620.3695022, 10.1145/3691620.3695024, 10.1145/3691620.3695055, 10.1145/3597503.3639103, 10.1145/3639476.3639777} \\
Dependency Version Issues & \cite{10.1145/3691620.3694982, 10.1145/3597503.3639103, 10.1145/3597503.3639194, 10.1145/3597503.3639184, 10.1145/3639476.3639762, 10.1145/3597503.3639183} \\
General Code Issues & \cite{10.1145/3691620.3694982, 10.1145/3691620.3695055, 10.1145/3597503.3608134, 10.1145/3639476.3639762, 10.1145/3597503.3639194} \\
Deprecated Models & \cite{10.1145/3597503.3623298, 10.1145/3597503.3639194, 10.1145/3597503.3608134, 10.1145/3639476.3639757, 10.1145/3597503.3639183} \\
Incomplete Documentation & \cite{10.1145/3691620.3695024, 10.1145/3597503.3608132, 10.1145/3597503.3639103, 10.1145/3597503.3639183} \\
Non-Executable Artefacts & \cite{10.1145/3597503.3623316, 10.1145/3597503.3639184} \\
Lacking Error Handling & \cite{10.1145/3639475.3640112, 10.1145/3639476.3639762} \\
Deprecated External API & \cite{10.1145/3597503.3623298} \\
\bottomrule
\end{tabular}
\label{tab:failure-reasons}
\end{table}

\begin{tcolorbox}[title=RQ2 Summary]
The main factors impeding the reproduction of \acrshort{llm}-centric empirical SE studies are missing or incomplete artefacts, missing details in reporting, dependency version issues, and, the only commercial LLM-specific factor: deprecated models. %Rev12
\end{tcolorbox}

\subsection{RQ3: \RQThree}
In total, of the analysed 85 articles, 18 were assigned an ACM badge. Among those articles, we found two types of badges: \textit{Artifacts Available v1.1} and the \textit{Artifacts Evaluated - Reusable v1.1}. Of the \acrshort{icse} papers, all 11 had the available and the reusable badge. At \acrshort{ase}, seven papers had the available badge, and 4 the reusable badge.  None of the articles had an \textit{Artifacts Evaluated - Functional v1.1} badge, which is the precursor to the reusable badge \cite{acm2020artifactreview}.
Of the three articles with only the \textit{Artifacts Available v1.1} badge, one resource did not provide the artefact anymore, as the repository expired, and one article was incomplete. Those results are in tune with ACM's description of this badge. ACM does not mandate completeness, nor a specific repository provider. Furthermore, ACM only explicitly mandates a permanent accessibility plan for data, ignoring code as a research artefact. \\
Of the 16 papers that were assigned the \textit{Artifacts Evaluated - Reusable v1.1} badge, four articles provided incomplete artefacts, three were non-functional, and one did not provide sufficient documentation for reproduction.
Those results contradict ACM's requirements for this badge. ACM mandates completeness, exerisability, consistency, and documentation to the level that it facilitates the reuse and repurposing of research artefacts. This is clearly not the case for half of the papers with the \textit{Artifacts Evaluated - Reusable v1.1} badge.

\begin{tcolorbox}[title=RQ3 Summary]
One year after the artefacts have been evaluated and awarded with the \textit{Artifacts Evaluated - Reusable v1.1} badges, half of them do not fulfill (anymore) the requirements set up by ACM. This suggests that ACM artefact badges are not a reliable indicator of reproducibility.
\end{tcolorbox}

\section{Discussion} \label{sec:discussion}
In this section, we critically reflect on our results and their implications for \acrshort{llm}-centric empirical studies. Afterwards, we provide recommendations for authors, venues, and funding agencies.
%%%%%% Findings %%%%%%
\subsection{Critical Reflection}
In the following, we discuss the factors coinciding with enhanced and degraded reproducibility, the value of LLM-centric empirical studies in light of our results, why to perform a reproduction study despite constant evolution of the technology under investigation, and the relevance of ACM artefact badges for the reproducibility of \acrshort{llm}-centric empirical research. 
\paragraph{Which factors influence reproducibility of \acrshort{llm}-centric studies?}
Intuitively, certain characteristics of \acrshort{llm} studies should coincide with enhanced reproducibility. Examples would be an ACM artefact evaluated badge, the minor version of the \acrshort{llm} being fixed, comparison with multiple models or metrics that are developed to evaluate \acrshort{llms} (e.g., pass@k). However, during our attempts of reproduction, we could not find any of those characteristics to be related to improved reproducibility, as reported in \Cref{tab:repl_factors}. However, from the factors impeding replication (see \Cref{tab:failure-reasons}), we can deduce the factors influencing degraded reproducibility. Those factors are most often not linked to \acrshort{llms}, but general issues in reproducibility that are discussed in research already, such as incomplete artefacts~\cite{10.1145/3477535, GONZALEZBARAHONA2023107318}, incomplete documentation~\cite{10.1007/s10664-011-9181-9}, or general code issues~\cite{Trisovic2022, HASSAN2025112327}. We will not discuss those factors in further detail, as the research community already offers a plethora of advice here (e.g., \cite{10.1007/s10664-021-09973-5, FRATTINI2024112120}). 
This highlights, that the reproducibility of LLM-centric studies largely suffers from the same issues as other empirical software engineering studies. %Rev12
The only factor directly linked to \acrshort{llms} was the usage of deprecated models, an issue exclusive to commercial models. We acknowledge that authors have little control over the deprecation of models. However, the risk of model deprecation has to be a consideration during the study design phase. %Rev18
Following the definition of the open source AI by the open-source initiative \cite{osi_open_source_ai_definition}, the hurdles towards deprecation for open source models are higher, and such models do not frequently get discontinued as opposed to commercial models. One way of dealing with model deprecation of commercial providers could hence be the comparison with open-source models. Four of the five studies we reviewed compared different models. However, analysing whether this comparison actually improves replicability is out of scope of this article. Nevertheless, on a bigger picture, half of the 85 articles did not perform a comparison with open source models, potentially risking model accessibility issues in the long term. \\
In the background, we mentioned the promise of commercial model providers, such as OpenAI, to keep models of a fixed version stable. It would have been an interesting analysis, whether this promise holds true. However, we can not make any claims given the few approaches that actually set the minor version for the used models (see \Cref{tab:repl_factors}).

\begin{table}[htp]
    \centering
    \caption{Characteristics of Reproduced Studies.}
    \begin{tabular}{@{}p{0.4\columnwidth}ccccc@{}}
        \toprule
        & \cite{10.1145/3597503.3639226} & \cite{10.1145/3597503.3639183} & \cite{10.1145/3639476.3639757} & \cite{10.1145/3639476.3639762} & \cite{10.1145/3691620.3694994} \\
        \midrule
        Reproducible & No & No & No & Part & Part \\
        \midrule
        ACM Artefact Evaluated & Yes & No & Yes & No & Yes \\
        Minor Version Set & No & Yes & No & No & Yes \\
        Model Comparison & Yes & Yes & Yes & Yes & No \\
        Metric for \acrshort{llms} & No & No & Yes & No & No \\
        \bottomrule
    \end{tabular}
    \label{tab:repl_factors}
\end{table}

\paragraph{The value of LLM-centric empirical studies}
Given that we started our replication efforts with 69 articles describing \acrshort{llm}-centric empirical research using OpenAI services, a partial reproduction of only two articles is alarming. More alarming than the low number of replications is the low number of executable studies, i.e., studies for which we could run the code successfully. With only five studies, this leaves 64 studies in a non-reproducible state. We could argue that those studies can still provide tangible value, e.g., by giving general guidance or by reporting on some key insights by the authors; this goes, however, under the assumption that the authors would describe their study setup in sufficient detail which is not the case. Of all studies, eight failed to even name the employed models, simply referring to \textit{ChatGPT} or to \textit{LLM}. Only a third of all studies reported on the temperature of their models and even less (16\%) reported on evaluation parameters such as \textit{top-p} or \textit{top-k}. Furthermore, only one in four studies discussed the aspect of context window size, not only hindering reproduction, but also demonstrating a large threat to validity to the available study results. Conversely, more than half of all studies reported on the employed prompts, contributing at least in part to their replicability.
In terms of reproducibility, \acrshort{llm}-centric empirical \acrshort{se} studies show similar issues as qualitative studies.
However, the difference is that for qualitative studies, we will rarely be able to reproduce (or simulate) the exact context with the same subject configuration as the baseline studies; this restriction, we argue, does not hold for \acrshort{llm}-centric empirical studies.
Therefore, we can only speculate that researchers need significantly more guidance.
This is especially important when---as a research community---we aim and claim to set the same standards with respect to scientific rigour and relevance for \acrshort{llm}-centric empirical studies as we do for other studies.

\paragraph{ACM Badges and Reproducibility}
Our findings raise questions about the long-term validity and relevance of the ACM artefact badges. Just one year after the evaluation, half of the artefacts with an \textit{Artifacts Evaluated - Resuable v1.1} badge failed to meet ACM requirements, rendering its long-term validity debatable. Given that the review processes are not standardised and no insights on the individual evaluations are available, we can not clearly attribute the observed discrepancies to the review process or to changes to the research artefacts over time, which would not be in scope of the awarded artefact badge. We can further operate under the assumption of appropriate initial artefact evaluation processes. At the same time, a preliminary analysis of artefacts showed that most of the reviewed artefacts have not been changed since the conferences in 2024. Of course, only findings are only representative for ICSE and ASE 2024, however given their role in the \acrshort{se} community, we are confident to %Rev7
argue for the need for a more standardised and transparent review process to improve the overall reproducibility of studies. So far, this does not seem to be intended by ACM who ''[...] believe that it is still too early to establish more specific guidelines for artifact and replicability review''~\cite{acm2020artifactreview}. \\
As the review process of artefacts is resource-intensive, one way to reduce reviewer effort would be an automatic pre-screening of artefacts. We do not go as far as to suggest the automation of the replication process itself, as even OpenAI admitted that \acrshort{llms} do not excel at that task yet \cite{starace2025paperbenchevaluatingaisability}. However, a source of inspiration could be the recently introduced PaperCheck \cite{papercheck}, which offers the tooling to scan scientific publications for adherence to best practices. A similar approach for research artefacts could provide both authors and reviewers with low-effort insights into improvement potentials.

\paragraph{The Value of a Reproducibility Study in a Fast-Paced Research Field}
Finally, one question we deem relevant is to which extent a reproducibility study in such a fast-paced research field can provide valuable insights (also considering the blurry notion of what can be seen as valuable).
As we discussed in the background, new models might improve overall on the benchmarks defined by the model vendors, but it is unclear how those updates impact individual---by the research community---defined tasks. We are not reproducing the results of model vendors on their benchmarks, but the model performance on tasks defined by the research community. If the models improved on all those tasks, we could see a reproducible value in the currently published studies as a lower bound to the potential of \acrshort{llms}. However, as our results for \cite{10.1145/3597503.3639183} and \cite{10.1145/3597503.3639226} indicate, this is not the case. \\
The reproduced studies were published just one year ago; hence, they represent the current state of knowledge. Despite the progress in the research field of AI focusing on LLMs, the general insights should remain the same. 
If the results of studies that are just one year old no longer hold true, we must question the sustainability of such research.
This is even more critical when considering that studies presented at conferences are typically part of larger research programmes contributing to the body of knowledge long-term, and often include contributions of individual PhD endeavours.

%%%%%% Recommendations %%%%%%
\subsection{Recommendations}
In the following, we provide recommendations for authors, venues, and funding agencies to improve the reproducibility of \acrshort{llm}-centric empirical studies.
\paragraph{Recommendations for Authors}
In 2024, \textcite{11071451} highlighted the need for guidance on the reporting of \acrshort{llm}-centric empirical studies. Addressing this gap, Baltes et al. published comprehensive guidelines in 2025 \cite{baltes2025guidelinesempiricalstudiessoftware}. For software engineering researchers at the intersection with other fields, it might prove valuable to check for additional guidance in those fields. For instance, the healthcare field addressed the need for reporting guidance in their TRIPOD-LLM Guidelines \cite{Gallifant2025TRIPODLLM}. As there exists extensive guidance on the planning, execution, and reporting of \acrshort{llm}-centric empirical studies, covering many of the challenges we reported in this paper, we will not provide further recommendations.

\paragraph{Recommendations for Venues}
Over the last years, we have seen venues including both, conferences and journals putting more emphasis on research artefacts. However, in light of the challenges we encountered, we suggest not only incentivising the disclosure of the full research artefacts, but also working towards a state where the disclosure is mandatory within the boundaries of what is possible (e.g., considering work rooted in sensitive data). This increased transparency must not only include code and data, but also a complete Software Bill of Materials. This recommendation is consistent with prevailing research on reproducibility. As discussed, the requirements for artefact evaluation laid out by ACM do not seem to suffice to ensure long-term reproducibility. Hence, we, as a research community, should rethink the criteria laid out and the guidance given towards artefact evaluation. This goes hand in hand with the strong recommendation to set the same standards for scientific rigour, reproducibility, and replicability as for other study types. We cannot continue to treat \acrshort{llm}-centric studies separately, on the basis of it being a ''new'' technology.

\paragraph{Recommendations for Funding Agencies}
In the same spirit as we make our recommendations for the future of our venues, we also expect funding agencies to have the same scientific rigour requirements for \acrshort{llm}-centric research as for other research. We believe that funding agencies can shape the long-term value of \acrshort{llm}-centric research by putting an emphasis on long-term accessibility, extending their mandatory practices towards open science, also to reproducibility, and by supporting the research community to improve the surrounding processes. One step towards such an improvement could be providing automation support in the assessment of research artefacts to lower the effort for reviewers and, thus, enabling higher quality outcomes in artefact disclosure and the peer review processes.

\section{Threats to Validity} \label{sec:threats}
We discuss the threats to validity of this study in terms of internal validity, external validity, and reliability, following the guidelines of \textcite{Runeson2009}. \\

\textit{Internal Validity}
First, we can only draw statistically significant conclusions if the sample size adequately accounts for the non-determinism of \acrshort{llms}. As discussed in \Cref{sec:background}, this was not feasible, and we resorted to 30 repetitions as an approximation to the real distribution.
Second, most research artefacts required manual intervention. Those interventions ranged from recovering dependency versions to code reconstructions, each with its own risk of introducing unintentional researcher bias. We minimized this threat by only reconstructing if the necessary documentation was available and additionally verified---where possible, the functionality with data provided in the research artefact.
Finally, the Bayesian bootstrap only considers observed values, potentially misrepresenting extreme behaviours of \acrshort{llms}.
In our experiments, this did not emerge as an issue. Nevertheless, we performed the descriptive analysis and provided all results transparently in our online material \cite{online-material}.

\textit{External Validity}
Several factors in our sampling process constrain the generalisability of our study.

First, we included only studies utilising OpenAI LLMs, thereby limiting applicability to models from other providers (e.g., Anthropic, Google, Meta), which may differ in architecture, availability, or versioning strategies.

Second, we only sampled studies from two conference proceedings, where the low number of successful study reproductions can not be generalised to other artefacts published in different venues.

Third, budgetary constraints limited the number of tasks reproduced per study. Alternative task selections could yield different performance outcomes, further impacting generalisability.

\textit{Reliability}
As we saw in \Cref{tab:all_ranges}, reproductions could result in high variability between different repetitions with up to 30\% differences in metrics. Those variabilities indicate that a larger sample size might provide more reliable findings. At the same time, this variability demonstrates the overall difficulties in reproducing the original values.

The reproduction of relevant studies was assigned to five of the authors, with each researcher conducting the replication individually, potentially introducing a researcher bias. To mitigate this bias, another author independently reviewed the results before finalising the conclusions on the specific research artefact.

The literature search process was adjusted due to the low number of suitable articles from ICSE and efficiency reasons. While this represents a deviation from the original research plan, its adaption was necessary to increase the sample size from 13 ICSE articles to 18 ICSE and ASE articles for reproduction attempts.
While the exclusion of articles from ASE without an artefact badge provides a drastic reduction, it contributed to the feasibility of the research within the allocated resources while strengthening the sample size from 11 ICSE articles to 18 ICSE and ASE articles for the ACM artefact badge analysis thus increasing the reliability of the results of RQ3.

\section{Conclusion} \label{sec:conclusion}
In this study, we investigated the reproducibility of \acrshort{llm}-centric empirical \acrshort{se} studies involving commercial \acrfull{llms}. We performed a critical review combined with a reproduction study. To that end, we reviewed 85 articles from \acrshort{icse} and \acrshort{ase} 2024, identifying 69 articles that used OpenAI's services. Of those, 18 articles included research artefacts and appeared suitable for reproduction. Ultimately, only five studies were fit for reproduction. Through our self-developed reproduction framework, we reproduced each of those five studies with up to 30 repetitions. 

Our findings reveal that none of the five studies could be fully reproduced. Two were partially reproducible, and three yielded divergent results. The most common reproduction challenges were missing or incomplete artefacts, dependency version issues, general code issues, and deprecated model access. We could--due to the low number of reproducible studies--not identify factors positively influencing reproducibility.
Even the presence of ACM artefact badges--intended to signal reliability--was no indicator of actual reproducibility.

These results suggest that despite a growing interest in \acrshort{llm}-based research in \acrshort{se}, the research field suffers from fundamental issues with reproducibility, similar to qualitative studies. This undermines both the scientific value of current \acrshort{llm}-based publications and the long-term relevance of their findings.

To address this, we offer recommendations for authors, venues, and funding agencies. We encourage authors to rigorously document all experimental data and share complete, versioned artefacts, aligned with relevant guidelines for the respective research domains. For venues, we advocate an independent reassessment of artefact evaluation criteria--beyond ACM's current standards--and a stronger emphasis on promoting artefact sharing. Finally, we believe that funding agencies can contribute to improving \acrshort{llm}-based research by prioritising reproducibility and long-term accessibility of research artefacts.

As the \acrshort{se} research field continues to evolve rapidly through \acrshort{llms}, reproducibility must not be treated as optional. We must put the same standards on \acrshort{llm}-centric research, as we claim for other research types. Without structural improvements, we risk building a body of knowledge that cannot be tested and consequently not be trusted.

\begin{acks}
This work was funded by the KKS foundation through the SERT Research Profile project (research profile grant 2018/010) at Blekinge Institute of Technology.
\end{acks}

\section*{Data availability }
The datasets generated during this study are published at: \cite{online-material}.
The reproduction framework is available at \cite{replikit}. 

\printbibliography

\end{document}